\documentclass[reprint,twocolumn]{revtex4}
\usepackage{amsfonts}
\usepackage{amsmath}
\usepackage{amssymb}
\usepackage{charter}
\usepackage{graphicx}
\usepackage{float}

\UseRawInputEncoding
\begin{document}

\title{Estimating IF shifts based on SU(1,1) interferometer }
\author{Yuetao Chen$^{1}$}
\author{Gaiqing Chen $^{1}$}
\author{Mengmeng Luo$^{2}$}
\author{Shoukang Chang$^{1}$}
\thanks{Corresponding author. changshoukang@stu.xjtu.edu.cn}
\author{Shaoyan Gao$^{1}$}
\thanks{Corresponding author. gaosy@xjtu.edu.cn}
\affiliation{$^{{\small 1}}$\textit{MOE Key Laboratory for Nonequilibrium Synthesis and
Modulation of Condensed Matter, Shaanxi Province Key Laboratory of Quantum
Information and Quantum Optoelectronic Devices, School of Physics, Xi'an
Jiaotong University, 710049, People's Republic of China}\\
$^{{\small 2}}$\textit{Department of Physics, Xi'an Jiaotong University City
College, Xi'an 710018, China}}

\begin{abstract}
IF (Imbert--Fedorov) shifts which refers to a transverse micro-displacement
occurs at the interface between two media. The estimation of such
micro-displacement enables a deeper understanding of light-matter
interactions. In this paper, we propose a theoretical scheme to investigate
the IF shifts and incident angle sensitivity by introducing SPR sensor into
the SU(1,1) interferometer. By injecting two coherent states in the SU(1,1)
interferometer, we obtain the sensitivity of the IF shifts and incident
angle based on the homodyne detection. Our results demonstrate that it is
possible to get the maximal IF shift and the optimal IF shifts sensitivity
simultaneously. Meanwhile, the orbit angular momentum carried by
Laguerre-Gauss (LG) beam is unfavorable for improving the IF shift
sensitivity. Furthermore, we have investigated the sensitivity of the
incident angle in our scheme and found that it is capable of surpassing the
sensitivity limit of\textbf{\ }$(6\times 10^{-6})%
{{}^\circ}%
.$ This allows us to achieve a more precise IF shifts sensitivity than the
traditional weak measurement method used for IF shift detection, which
typically has a rotation precision limit of 0.04$%
{{}^\circ}%
$ [Journal of Optics, 19(10), 105611]. More importantly, both the
sensitivity of IF shifts and incident angle can breakthrough the (shot noise
limit) SNL, even approaching the Cram\'{e}r-Rao bound (QCRB) at the incident
angle $\theta =43.6208%
{{}^\circ}%
$ and $\theta =43.6407%
{{}^\circ}%
$. We also discover that increasing the coherent amplitude is beneficial for
improving the sensitivity of both the IF shifts and incident angle. Our
findings shall offer a novel scheme for measuring micro-displacement in SPR
sensor. These results can be helpful in the development of more precise
quantum-based sensors for studying light-matter interactions.

\end{abstract}

\maketitle

\section{Introduction}

Quantum metrology has attracted wide attention since it can achieve
ultra-high precision estimation of various physical quantities by using some
quantum effects, such as squeezing, nonclassicality and entanglement, and
therefore has a lot of applications in many other fields, including
gravitational wave detection \cite{7,8}, quantum imaging \cite{9,10,11} and
quantum thermometry \cite{12,13,14}. As one of the important platforms to
realize quantum metrology, optical interferometer has attracted extensive
attention in recent years. Accordingly, various optical interferometers have
been proposed to realize the precise measurement of many physical
quantities, such as phase shift, angular displacement and field-quadrature
displacement. At the beginning of quantum metrology, one mainly injects some
non-classical quantum states into the traditional Mach--Zehnder
interferometer (MZI) to enhance the phase or angular displacement
sensitivity. In particular, Caves proposed a feasible scheme, that is, using
squeezed vacuum state instead of coherent state to significantly improve the
phase sensitivity of MZI. Since then, non-classical quantum states,
involving twin Fock states, NOON state and two-mode squeezed vacuum state,
are used for achieving higher phase shift or angular displacement precision,
even beating the Heisenberg limit (HL). Although using non-classical states
can realize the improvement of phase or angular displacement sensitivity of
the optical interferometer, these quantum states are not only difficult to
generate by using existing experimental techniques, but also very vulnerable
to the impact of noise environment. Hence one began to pay attention to
whether the devices of optical interferometer can enhance the measurement
accuracy. For instance, Yurke \emph{et al}. suggested for the first time to
replace two linear optical beam splitters of the conventional MZI with
nonlinear optical parametric amplifiers, thus putting forward the SU(1,1)
interferometer and further showed that the phase sensitivity of the SU(1,1)
interferometer can break through the shot noise limit (SNL). In recent
decades, many theoretical and experimental schemes based on the SU(1,1)
interferometer have been proposed extensively. One of these schemes was
given by Li \emph{et al}., who theoretically demonstrated that the SU(1,1)
interferometer based on the homodyne detection can realize the HL
sensitivity by using a coherent state and a squeezed vacuum state as input
states. Subsequently, Liu \emph{et al}. showed that the SNL of the angular
displacement can be surpassed by using a coherent state as the input state
of the SU(1,1) interferometer.

In addition to the phase estimation, angular displacement estimation and
field-quadrature displacement estimation, the quantum sensor based on the
Kretschmann structure has attracted wide attention in experiment and theory.
Especially, the two types of nonlinear interferometric
surface-plasmon-resonance sensor proposed by Wang \emph{et al}., in which
Kretschmann structure is placed inside or outside SU(1,1) interferometer
\cite{43}. We also noticed that the Imbert--Fedorov (IF) shift estimation
and incident angle estimation has not been studied yet based on the SU(1,1)
interferometer. It is well known that when\ a bounded light beam is
illuminated to an interface of two different medium, the reflected light
beam will experience transverse shifts which are called IF shifts \cite{26}.
IF shifts depend sensitively on the refractive index of the medium. In other
words, the information of refractive index of the medium can be discerned by
observing the variation in IF shifts. Thus, IF shifts attract renewed
attention in various field such as optical\ sensors \cite{27}, precision
metrology and nanophotonic devices \cite{28,29,30}. Generally, large IF
shifts can be obtained in the surface plasmon resonance (SPR) sensors and
measured in weak measurement scheme.\textbf{\ }However, there are always
existing error in SPR sensors which will lead to low precision in the
measurement of IF shifts.\textbf{\ }Besides, in recent decades, it is found
that no matter how to improve the traditional measurement scheme by using
traditional means, the measurement accuracy can not be substantially
improved owing to the existence of classical noise. The precision limit
constrained by classical noise is known as the shot-noise limit (also known
as the standard quantum limit)\textbf{\ }\cite{31,32}\textbf{. }As described
above, the SU(1,1) interferometer which is an important tool for the
precision measurement could also be good candidates for the IF shifts
measurement. Thus, in this paper, we theoretically investigate the IF shift
sensitivity for a Laguerre Gaussian reflected beam in a surface plasmon
resonance sensor via the SU(1,1) interferometer. By taking the homodyne
detection at one of output ports of the SU(1,1) interferometer, we obtain
the IF shift and incident angle sensitivity. Then, the SNL and the quantum
Cram\'{e}r-Rao bound (QCRB) are also analytically derived. The numerical
results show that the sensitivity of IF shifts and incident angle can
surpass the SNL, even approaching\textbf{\ }the QCRB\textbf{\ }around the
SPR angle. Moreover, the maximal IF shifts and the minimum IF shifts
sensitivity can be obtained simultaneously. The incident angle sensitivitiy
can break through $(6\times 10^{-6})%
{{}^\circ}%
$ which shows more accurate compared with the rotation precision of 0.04$%
{{}^\circ}%
$ in the weak measurement of IF shifts \cite{34}.

The paper is organized as follows. In Sec. II, we present a description of
the Kretschmann structure and IF shifts. In Sec. III, we describe the
interaction process between incident coherent beam and the SPR sensors in
nonlinear transmissivity estimation model. In Sec. IV, the sensitivity of IF
shifts and incident angle is investigated via homodyne detection. In Sec. V,
we obtain the QCRB and SNL of IF shifts and incident angle and compare them
with the sensitivity of IF shifts and incident angle. Finally, the main
conclusions are presented in the last section.

\section{Kretschmann structure and Imbert--Fedorov shifts}

For the sake of facilitating the discussion of the Imbert--Fedorov shift
estimation based on the SU(1,1) interferometer, we briefly review the known
results of the Kretschmann structure and the Imbert--Fedorov (IF) shifts in
this section. As shown in Fig. 1, the Laguerre--Gaussian beam is illuminated
to the Kretschmann structure which consists of three layers, the top layer
is a prism, the middle is a thin gold film coated on prism which leads to
surface plasmon resonance and the bottom is semi-infinite vacuum. The global
coordinate system is marked as $(x_{g};y_{g};z_{g})$, while the local
coordinate systems are represented by $(x_{i};y_{i};z_{i})$ and $%
(x_{r};y_{r};z_{r})$ which are attached to the incident and reflected beams,
respectively. Assume that a TM-polarized Laguerre-Gauss beam is incident
onto the Au film from the glass prism, after reflection there will be IF
shifts in the $y_{g}$ direction at the glass-Au interface. The angular
spectrum of the incident TM-polarized Laguerre-Gauss beam can be represented
by
\begin{equation}
\overset{\sim }{E_{i}}=[w_{0}(-ik_{ix}+\text{sgn}[l])k_{iy}/\sqrt{2}%
]^{\left\vert l\right\vert }e^{-(k_{ix}^{2}+k_{iy}^{2})w_{0}^{2}/4},
\label{1}
\end{equation}%
where $l$ and $%
k_{i\lambda }$ $(\lambda =x,y)$ is the units of orbit angular momentum and
the lateral wave vectors of incident beam, respectively, the beam waist $%
w_{0}$ is 1000$\mu m,$ the symbolic function sgn$[l]$ can be defined by
\begin{equation}
\text{sgn}[l]=\left\{
\begin{array}{c}
1,l>0 \\
0,l=0 \\
-1,l<0%
\end{array}%
\right. .  \label{2}
\end{equation}%
The Imbert--Fedorov shifts are related to the complex reflection coefficient
$r_{pgv}$ of the Kretschmann configuration given by
\begin{equation}
r_{pgv}=\frac{r_{pg}+r_{gv}e^{2ik_{gz}d}}{1+r_{pg}r_{gv}e^{2ik_{gz}d}},\
\label{3}
\end{equation}%
where $d$ denotes the thickness of gold
film and the reflection coefficient between the $n$th and $m$th layers $%
r_{nm}$ can be expressed as
\begin{equation}
\ r_{nm}=\frac{k_{nz}\varepsilon _{m}-k_{mz}\varepsilon _{n}}{%
k_{nz}\varepsilon _{m}+k_{mz}\varepsilon _{n}},  \label{4}
\end{equation}%
where $n,m$=(p(prism), g(gold), and
v(vacuum)), and\ $k_{nz}$ represents the normal component of the wave vector
in the $n$th layer and can be given by%
\begin{eqnarray}
\ k_{nz} &=&\frac{2\pi }{\lambda }\sqrt{\varepsilon _{n}-\varepsilon
_{p}(\sin \theta )^{2}},\   \notag \\
k_{mz} &=&\frac{2\pi }{\lambda }\sqrt{\varepsilon _{m}-\varepsilon _{p}(\sin
\theta )^{2}},  \label{5}
\end{eqnarray}%
where $%
\theta $ is the incident angle, $\varepsilon _{p}$ and $\varepsilon _{g}$
are the permittivities of prism and gold film, respectively, $\lambda $ is
the wavelength of incident light of TM-polarized Laguerre-Gauss beam. For
simplicity, we choose the gold film with the thickness $d=47nm$, the prism
and vacuum are semi-infinite. In order to visually see what the value of the
incident angle takes, a large IF shifts can be generated. At fixed values of
$\varepsilon _{p}=2.22$, $\varepsilon _{g}=-20.327+1.862i$, $d=47nm$ and $%
\lambda =780$nm, we plot the the reflectivity $\left\vert r_{pgv}\right\vert
$ of Kretschmann structure as a function of the incident angle $¦È$
as shown in Fig. 1(b). It is clear that the reflectivity $\left\vert
r_{pgv}\right\vert $ can reach the minimum value at the incident angle $%
\theta =43.63^{\circ },$ and this incident angle is also called the SPR
angle. The reason for this phenomenon is that the excitation of surface
plasmon modes and reflectivity is dramatically attenuated at this incident
angle. The sharp variation in reflectivity near the SPR angle will give rise
to huge IF shifts and the shifts can be simplified as follow in the
condition of large\textbf{\ }$w_{0}$\textbf{\ }\cite{35}.
\begin{equation}
Y=-l\dfrac{\partial \left\vert r_{pgv}\right\vert /\partial \theta }{%
k_{0}\left\vert r_{pgv}\right\vert },  \label{6}
\end{equation}%
where $k_{0}$ is wave vector of incident
light in vacuum.

\begin{figure}[tbp]
\label{Fig1} \centering \includegraphics[width=0.75\columnwidth]{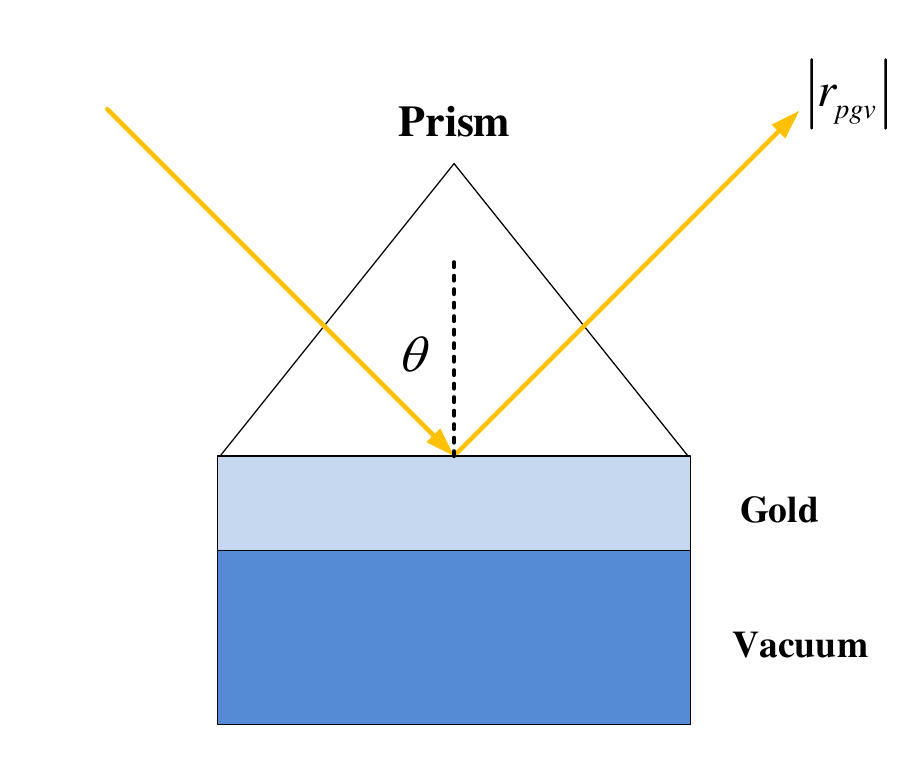}%
\newline
\includegraphics[width=0.75\columnwidth]{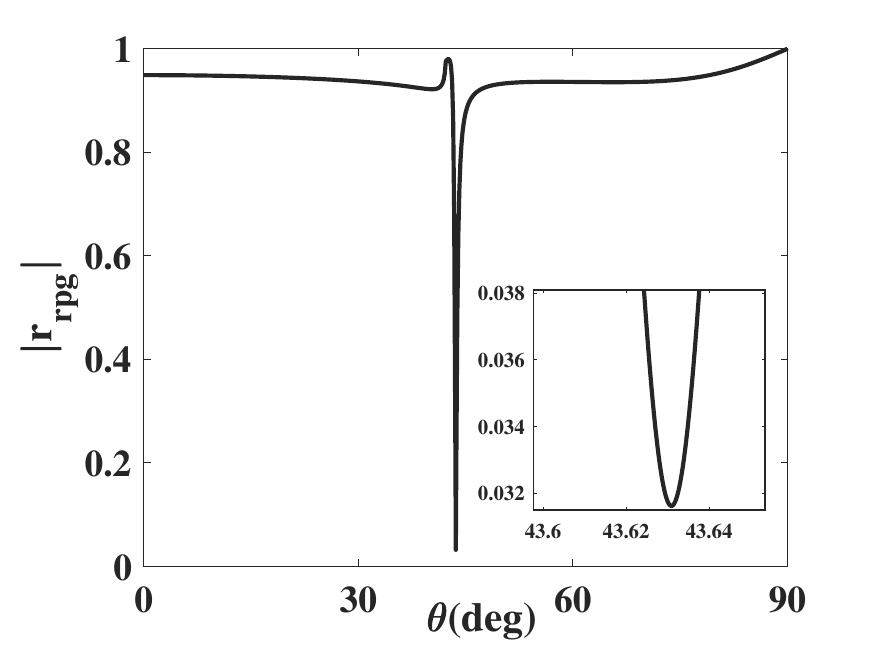}\newline
\caption{{}(Color online) (a)An attenuated total reflection for
Laguerre--Gaussian beam in the SPR sensor. (b)Reflectivity $\left\vert r_{pgv}\right\vert
$ of Kretschmann structure as a function of the incident angle $¦È$.}
\end{figure}
In Fig. 2, we plot the $Y/\lambda $\ as a function of the incident angle $%
¦È $, when given parameters $\varepsilon _{p}=2.22$, $\varepsilon
_{g}=-20.327+1.862i$, $d=47nm$\ and $\lambda =780$nm.
According to Eq. (\ref{6}), $Y$\ varies with $(\partial \left \vert
r_{pgv}\right \vert /\partial \theta )/\left \vert r_{pgv}\right \vert $,
therefore $Y$\ can be enhanced when the SPR is excited. At the SPR angle, $%
Y=0$\ since $\partial \left \vert r_{pgv}\right \vert /\partial \theta =0$.
When the incident angle $¦È $\ cross the SPR angle, the value of $Y$\
changes from positive to negative, as $\partial \left \vert
r_{pgv}\right
\vert /\partial \theta $\ changes sign. Thus, a positive peak
and a negative peak can be found in Fig. 2, which locate at $¦È =43.6208%
{{}^\circ}%
$\ and $¦È =43.6407%
{{}^\circ}%
$,\ respectively. For different incident OAM $l$, these two kinds of peaks
can always be exsiting, as shown by Fig. 2. As OAM $l$\ increase, IF shifts
will be amplified linearly with it according to Eq. (6) and the peak
positions are always located at the SPR angle. The maximum shifts for the
cases of $l$\ =$\pm 3$\ are 1092$\mu m$. Thus, huge IF shifts can be
obtained in Kretschmann structure with Laguerre--Gaussian incident beam.
This also means that we are able to capture the properties of medium such as
excitation of surface plasmon modes from IF shifts. However, there always
exist error in IF displacement due to the limitation of incident angle
accuracy. Therefore, we need device to get accurate measurement of
displacement.\textbf{\ }
\begin{figure}[tbp]
\label{Fig2} \centering \includegraphics[width=0.75\columnwidth]{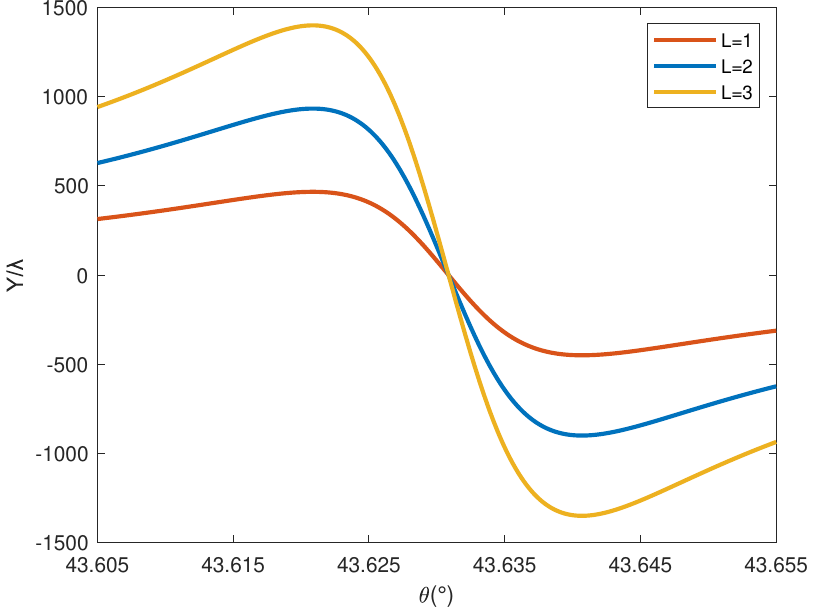}%
\newline
\newline
\caption{{}(Color online) IF shifts as a function of the incident angle $%
\protect\theta $ with $L=1$, $L=2$ and $L=3$.}
\end{figure}

\section{Nonlinear transmissivity estimation model}

The fundamental condition of the IF shift have been discussed in the
previous section, in what follows, we will introduce the IF shifts sensors
via the SU(1,1) interferometer. As shown in Fig. 3, we consider two coherent
states $\left\vert \alpha \right\rangle _{a}\otimes \left\vert \beta
\right\rangle _{b}$ with $z=\left\vert z\right\vert e^{i\theta _{z}},$ $%
z=\alpha ,\beta $ as the inputs in mode $a$ and mode $b$, respectively.
After going through the first OPA, mode $b$ is regarded as a reference beam,
while mode $a$ enters the Kretschmann structure to generate the IF shift to
be estimated. The action of the first OPA on two coherent states $\left\vert
\alpha \right\rangle _{a}\otimes \left\vert \beta \right\rangle _{b}$ can be
described by the unitary operator $\hat{S}(\xi _{1})=\exp (\xi _{1}^{\ast }%
\hat{a}\hat{b}-\xi _{1}\hat{a}^{\dagger }\hat{b}^{\dagger }),\xi
_{1}=g_{1}e^{i\theta _{1}}.$The interaction process between mode $a$ and the
Kretschmann structure can be simulated by the fictitious beam splitter $\hat{%
U}_{BS}=\exp [(\hat{a}^{\dagger }\hat{a}_{\upsilon }-\hat{a}\hat{a}%
_{\upsilon }^{\dagger })\arctan \sqrt{(1-\eta )/\eta }]$ with the
transmissivity $\eta =\left\vert r_{pgv}\right\vert ^{2}$, which can be
achieved experimentally by inserting the variable neutral density filter in
the optical path $a$ \cite{36}. Then, the modes $a$ and $b$ recombine in the
second OPA. Let $\hat{a}(\hat{a}^{\dagger })$ and $\hat{b}(\hat{b}^{\dagger
})$ are the annihilation (creation) operators for the modes $a$ and $b$,
respectively. Then, the relation between the output operators $\hat{a}_{2}$
and $\hat{b}_{2}$ and the input operators $\hat{a}_{0}$ and $\hat{b}_{0}$
can be expressed as

\begin{eqnarray}
\hat{a}_{2} &=&W_{1}\hat{a}_{0}-W_{2}\hat{b}_{0}^{\dagger }+W_{3}\hat{a}_{v},
\notag \\
\hat{b}_{2} &=&W_{4}\hat{b}_{0}-W_{5}a_{0}^{\dagger }+W_{6}\hat{a}%
_{v}^{\dagger },  \label{7}
\end{eqnarray}%
where
\begin{eqnarray}
W_{1} &=&\sqrt{\eta }\cosh g_{1}\cosh g_{2}+e^{i(\theta _{2}-\theta
_{1})}\sinh g_{1}\sinh g_{2},  \notag \\
W_{2} &=&\sqrt{\eta }e^{i\theta _{1}}\sinh g_{1}\cosh g_{2}+e^{i\theta
_{2}}\cosh g_{1}\sinh g_{2},  \notag \\
W_{3} &=&\sqrt{1-\eta }\cosh g_{2},  \notag \\
W_{4} &=&\cosh g_{1}\cosh g_{2}+\sqrt{\eta }e^{i\left( \theta _{2}-\theta
_{1}\right) }\sinh g_{1}\sinh g_{2},  \notag \\
W_{5} &=&e^{i\theta _{1}}\sinh g_{1}\cosh g_{2}+\sqrt{\eta }e^{i\theta
_{2}}\cosh g_{1}\sinh g_{2},  \notag \\
W_{6} &=&-\sqrt{1-\eta }e^{i\theta _{2}}\sinh g_{2},  \label{8}
\end{eqnarray}%
where $g_{j}$ and $\theta _{j}$ $(j=1,2)$ are the gain factor and phase
shift of OPA$_{j}$ process, respectively, and $\hat{a}_{v}$ correspond to
the input vacuum field associated with the introduction of the fictitious
beam splitter \cite{36}.

\begin{figure}[tbp]
\label{Fig3} \centering \includegraphics[width=1\columnwidth]{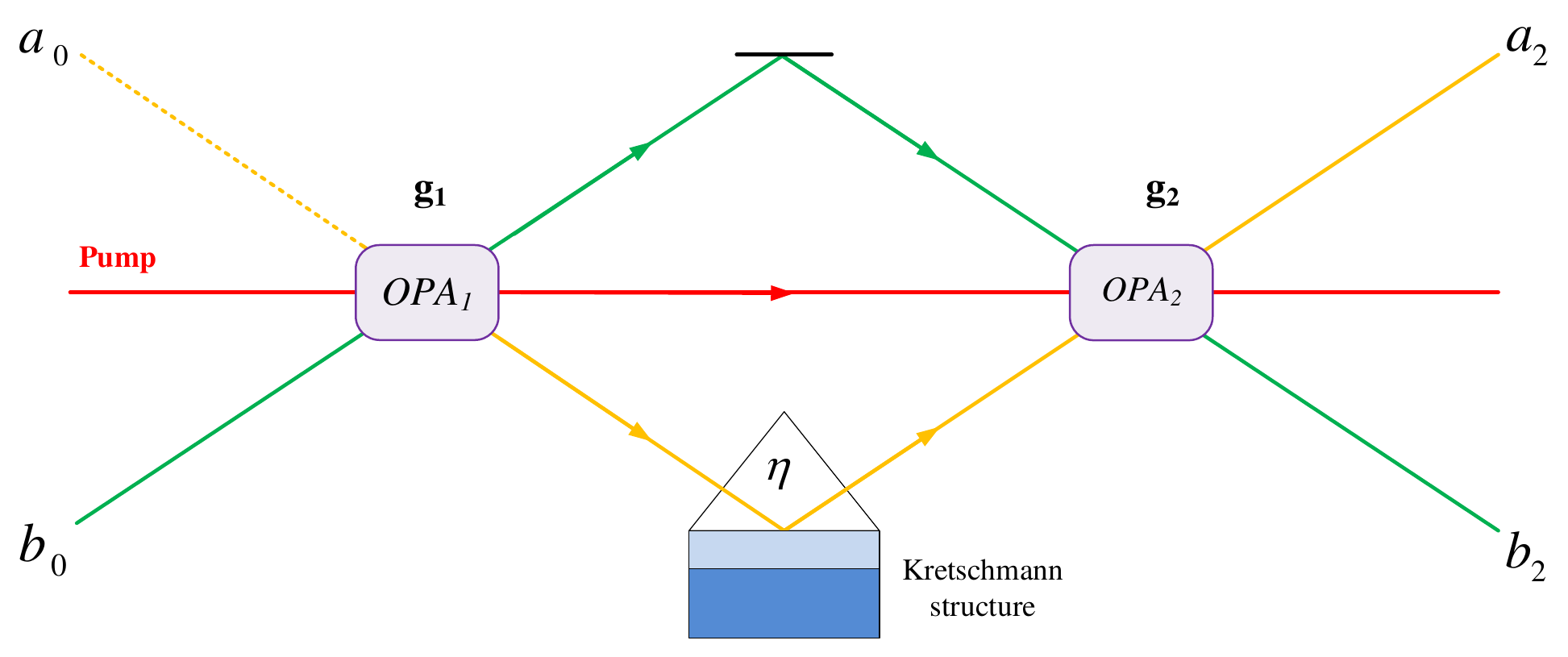}
\newline
\caption{{}(Color online) Schematic diagram of SU(1,1) interferometer with
SPR sensors. The two input ports of this interferometer are a coherent state
$\left\vert \protect\alpha \right\rangle _{a}$ and a vacuum state $%
\left\vert \protect\beta \right\rangle _{b}$}
\end{figure}

\section{IF shifts sensitivity via homodyne detection}

In this section, we mainly investigate the IF shift sensitivity based on the
SPR sensor. To this end, we need to choose a specific detection method for
reading the IF shift information at the final output port. In fact, there
are many distinct types of detecting methods, such as homodyne detection
\cite{24}, intensity detection \cite{37,38,39}, and parity detection \cite%
{40,41,42}. Moreover, compared with the intensity detection and parity
detection, homodyne detection is easily achieved using existing experimental
technology. Therefore, we use the homodyne detection to estimate the IF
shift, the corresponding detected variable is the amplitude quadrature $\hat{%
X}$, i.e.,

\begin{equation}
\ \hat{X}=\frac{\hat{a}_{2}+\hat{a}_{2}^{\dagger }}{\sqrt{2}}.  \label{9}
\end{equation}%
For the sake of discussion, we
consider the balanced situation between two OPAs processes, i.e., $\theta
_{2}-\theta _{1}=\pi ,$ and $g_{1}=g_{2}=g$ (let $\theta _{1}=0$ for
simplicity) in the following. According to the error propagation formula,
the IF shift sensitivity can be given by

\begin{eqnarray}
\Delta Y &=&\frac{\sqrt{\Delta ^{2}\hat{X}}}{\left \vert \partial \left
\langle \hat{X}\right \rangle /\partial Y\right \vert },  \notag \\
&=&\frac{\sqrt{\Delta ^{2}\hat{X}}}{\left \vert (\partial \left \langle \hat{%
X}\right \rangle /\partial \eta )(\partial \eta /\partial Y)\right \vert },
\label{10}
\end{eqnarray}%
where
\begin{eqnarray}
\Delta ^{2}\hat{X} &=&\left \langle \hat{X}^{2}\right \rangle -\left \langle
\hat{X}\right \rangle ^{2}  \notag \\
&=&\frac{1}{2}[(\sqrt{\eta }\cosh ^{2}g-\sinh ^{2}g)^{2}  \notag \\
&&+(\sqrt{\eta }-1)^{2}\sinh ^{2}g\cosh ^{2}g  \notag \\
&&+\left( 1-\eta \right) \cosh ^{2}g],  \notag \\
\left \langle \hat{X}\right \rangle &=&\sqrt{2}[\left \vert \alpha \right
\vert (\sqrt{\eta }\cosh ^{2}g-\sinh ^{2}g)\cos \theta _{\alpha }  \notag \\
&&-\left \vert \beta \right \vert (\sqrt{\eta }-1)\sinh g\cosh g\cos \theta
_{\beta }],  \notag \\
Y &=&-l\dfrac{\partial \eta /\partial \theta }{k_{0}\eta }.  \label{11}
\end{eqnarray}

\begin{figure}[tbp]
\label{Fig4} \centering \includegraphics[width=0.75\columnwidth]{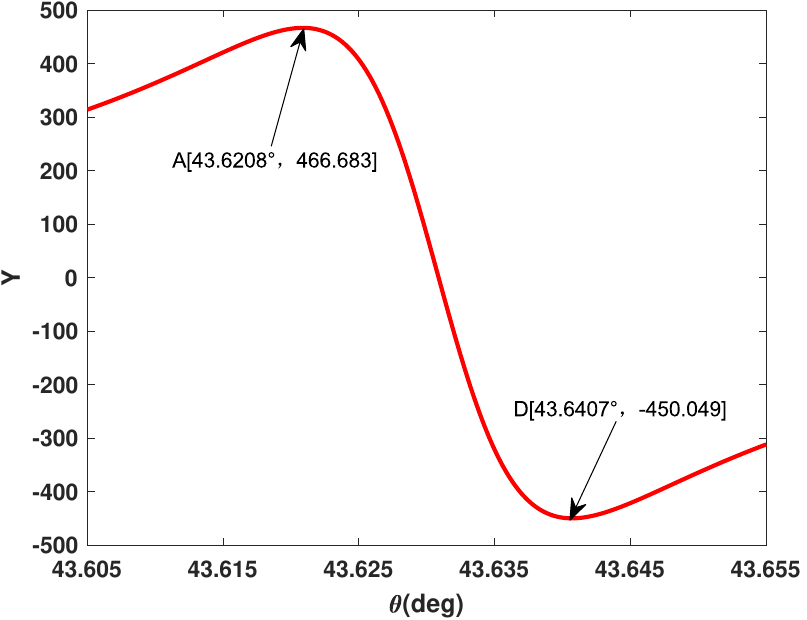}%
\newline
\includegraphics[width=0.75\columnwidth]{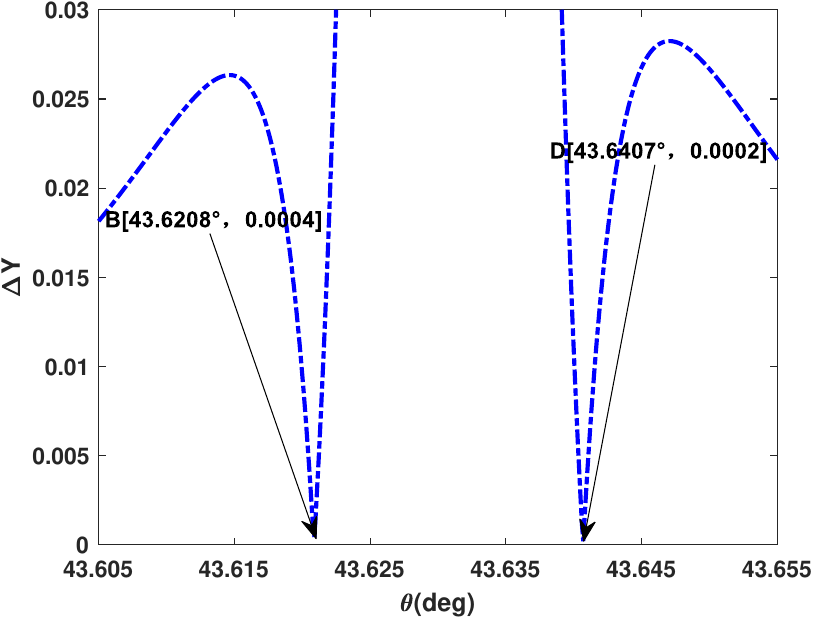} \newline
\caption{(a) IF shifts in SPR sensors, (b) IF shifts sensitivity based on
homodyne detection as a function of $\protect\theta $ with $g=0.7$, $\protect%
\theta _{\protect\alpha }=0$, $\protect\theta _{\protect\beta }=\protect\pi $%
, $L=1$, and $\left \vert \protect\alpha \right \vert =\left \vert \protect%
\beta \right \vert =50000$. The blue-dashed line and the black-solid line
correspond to IF shifts sensitivity ($\Delta _{Y}$) and IF shifts ($Y$),
respectively.}
\end{figure}

In Fig. 4(a), we show IF shifts $Y$ changing with $\theta $ for
Laguerre Gaussian reflected beam when given parameters $g=0.7$, $\theta
_{\alpha }=0$, $\theta _{\beta }=\pi $, $L=1$, and $\left\vert \alpha
\right\vert =\left\vert \beta \right\vert =50000$. We can clearly see that
there is a positive and negative peaks of IF shifts which is located at $%
\theta =43.6208%
{{}^\circ}%
$ and $\theta =43.6407%
{{}^\circ}%
$ on two sides of the SPR angle and IF shifts vanish at the SPR angle\textbf{%
\ (}$\theta =43.6309%
{{}^\circ}%
$\textbf{)}. The corresponding IF shifts sensitivity $\Delta Y$ as a
function of $\theta $ is shown in Fig. 4(b). It is clearly seen that IF
shifts sensitivity $\Delta Y$ experiences sharp variations at two dips
located at $\theta =43.6208%
{{}^\circ}%
$ and $\theta =43.6407%
{{}^\circ}%
$ on two sides of SPR angle and the minimum value of the IF shifts
sensitivity can be achieved at $\theta =43.6208%
{{}^\circ}%
$ and $\theta =43.6407%
{{}^\circ}%
$. This implies that large IF shifts and small IF shifts sensitivity can be
obtained simultaneously.\ Futher, to show how the IF shifts sensitivity is
affected by the coherence amplitude $\left\vert \alpha \right\vert $ and $%
\left\vert \beta \right\vert $, at $\theta =43.6208%
{{}^\circ}%
$ and $\theta =43.6407%
{{}^\circ}%
$, we also plot the IF shifts sensitivity as a function of $\left\vert
\alpha \right\vert $ and $\left\vert \beta \right\vert $ in Fig. 5(a) and
5(b), respectively. It can be found that, IF shifts sensitivity $\Delta Y$
decrease sharply with the increase of $\left\vert \alpha \right\vert $ and $%
\left\vert \beta \right\vert $, for the case of $\theta =43.6208%
{{}^\circ}%
$ and $\theta =43.6407%
{{}^\circ}%
$.

\begin{figure}[tbp]
\label{Fig5} \centering \includegraphics[width=0.75\columnwidth]{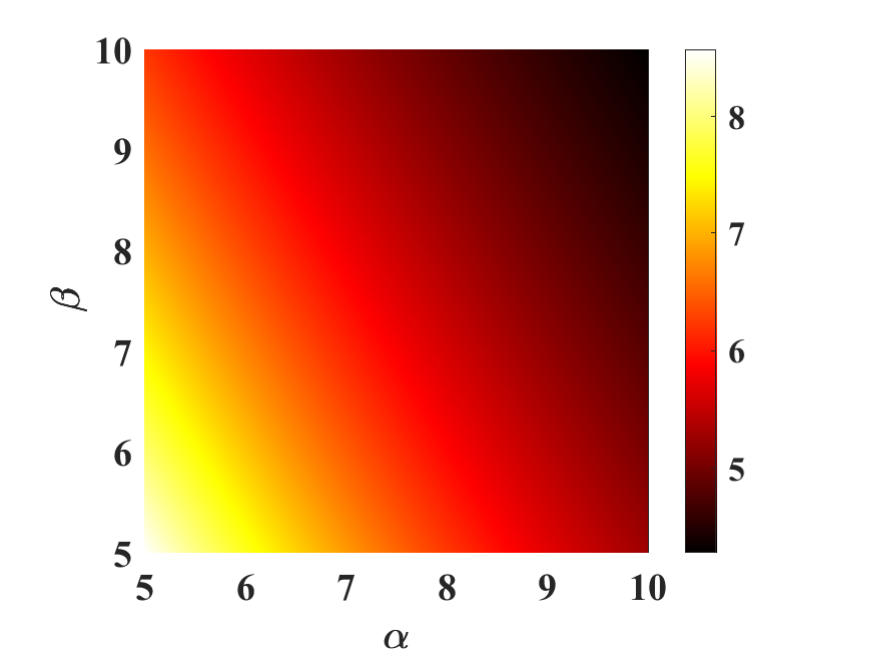}%
\newline
\includegraphics[width=0.75\columnwidth]{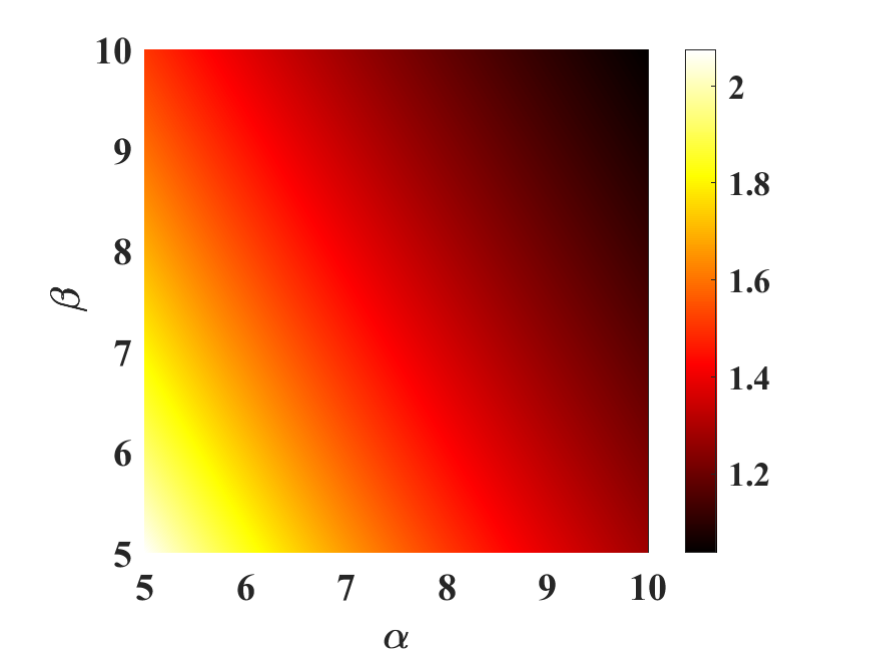} \newline
\caption{(Color online) IF shifts sensitivity based on homodyne detection as
a function of the coherent amplitude $\left\vert \protect\alpha \right\vert $
and $\left\vert \protect\beta \right\vert $ with $g=0.7$, $\protect\theta _{%
\protect\alpha }=0$, $\protect\theta _{\protect\beta }=\protect\pi $ and $%
L=1 $ at (a) $\protect\theta =43.6208%
{{}^\circ}%
$, and (b) $\protect\theta =43.6407%
{{}^\circ}%
$, respectively. }
\end{figure}

Moreover, the effect of the orbital angular momentum on IF shifts
sensitivity is also shown in Fig. 6 with different $L$\ changing with $%
\theta $ when other parameters are the same as those in Fig. 4.\ It is
obviously that the minimum value of the IF shifts sensitivity can still be
obtained at the same point $\theta =43.6208%
{{}^\circ}%
$ and $\theta =43.6407%
{{}^\circ}%
$ with different $L$, and the IF shifts sensitivity increases linearly with $%
L$. Thus, we conclude that large IF shifts and small IF shifts sensitivity
can be obtained at $\theta =43.6208%
{{}^\circ}%
$ and $\theta =43.6407%
{{}^\circ}%
$ simultaneously with different $L$.
\begin{figure}[tbp]
\label{Fig6} \centering \includegraphics[width=0.75\columnwidth]{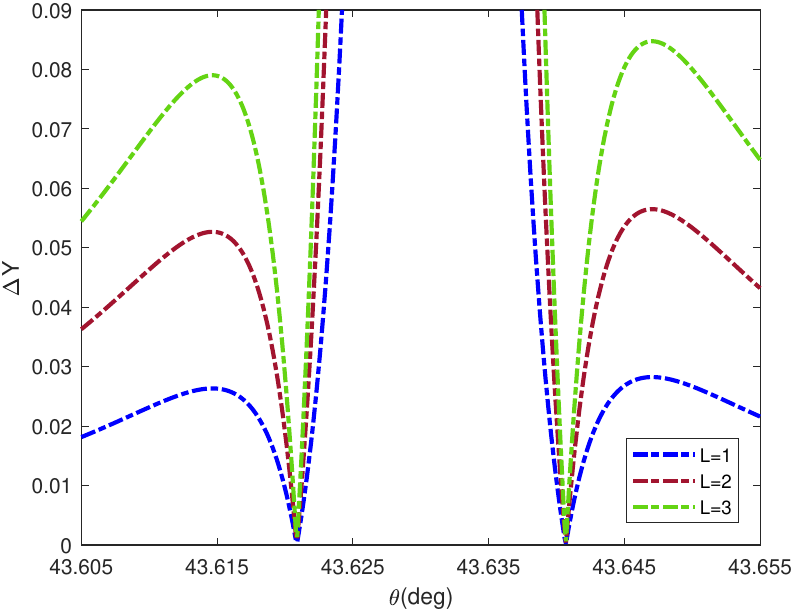}%
\newline
\newline
\caption{(Color online) IF shifts sensitivity based on homodyne detection as
a function of $\protect\theta $ with $g=0.7$, $\protect\theta _{\protect%
\alpha }=0$, $\protect\theta _{\protect\beta }=\protect\pi $, and $%
\left\vert \protect\alpha \right\vert =\left\vert \protect\beta \right\vert
=50000$ for different $L$. The blue-dashed line, the wine-red-dashed line
line and the green-dashed line correspond to $L=1$, $L=2$ and $L=3$,
respectively. }
\end{figure}

\section{Incident angle sensitivity via homodyne detection}

In the previous section, we have discussed the IF shifts sensitivity based
on the homodyne detection. It's worth noting that the IF shifts can be
observed in the weak measurement scheme \cite{34}. Moreover, the rotation of
the polarizer in the weak measurement scheme \cite{34} will lead to the
error of incident angle which is the main cause of the error of IF shifts ($%
\Delta Y$). For this reason, we further need to investigate the incident
angle sensitivity. In a similar way to obtain the IF shift sensitivity in
section 4, one can also derive\textbf{\ }the incident angle sensitivity

\begin{eqnarray}
\Delta \theta &=&\frac{\sqrt{\Delta ^{2}\hat{X}}}{\left\vert (\partial
\left\langle \hat{X}\right\rangle /\partial \eta )(\partial \eta /\partial
\theta )\right\vert },  \notag \\
\eta &=&\left\vert r_{pgv}\right\vert ^{2}=\left\vert \frac{%
r_{pg}+r_{gv}e^{2ik_{gz}d}}{1+r_{pg}r_{gv}e^{2ik_{gz}d}}\right\vert ^{2},
\notag \\
r_{nm} &=&\frac{k_{nz}\varepsilon _{m}-k_{mz}\varepsilon _{n}}{%
k_{nz}\varepsilon _{m}+k_{mz}\varepsilon _{n}},  \notag \\
k_{nz} &=&\frac{2\pi }{\lambda }\sqrt{\varepsilon _{n}-\varepsilon _{p}(\sin
\theta )^{2}},  \notag \\
k_{mz} &=&\frac{2\pi }{\lambda }\sqrt{\varepsilon _{m}-\varepsilon _{p}(\sin
\theta )^{2}},  \label{12}
\end{eqnarray}%
where $\Delta ^{2}\hat{X}$ and $\left\langle \hat{X}\right\rangle $ can be
obtained from Eq. (\ref{11}).

In Fig. 7, we show that the incident angle sensitivity $\Delta \theta $
changing with $\theta $ when given parameters $g=0.7$, $\theta _{\alpha }=0$%
, $\theta _{\beta }=\pi $ and $\left\vert \alpha \right\vert =\left\vert
\beta \right\vert =50000$. It can be found that $\Delta \theta $ is
relatively small around $\theta =43.6208%
{{}^\circ}%
$ and $\theta =43.6407%
{{}^\circ}%
$ except for the SPR angle. It is worth noting that the incident angle
sensitivitiy can break through $(6\times 10^{-6})%
{{}^\circ}%
$ which shows more accurate compared with the rotation precision of 0.04$%
{{}^\circ}%
$ in the weak measurement of IF shifts \cite{34}.

\begin{figure}[tbp]
\label{Fig7} \centering \includegraphics[width=0.75\columnwidth]{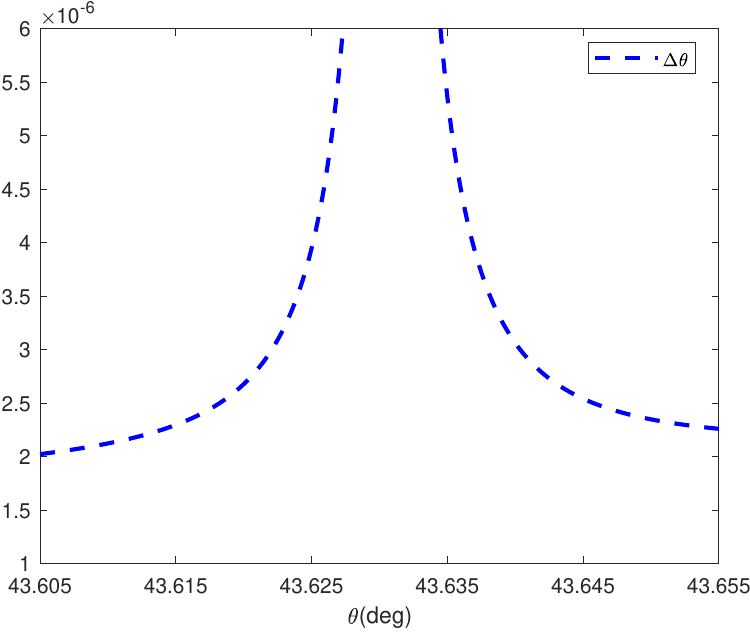}%
\newline
\newline
\caption{(Color online) Incident angle sensitivity based on homodyne
detection as a function of $\protect\theta $ with $g=0.7$, $\protect\theta _{%
\protect\alpha }=0$, $\protect\theta _{\protect\beta }=\protect\pi $, and $%
\left\vert \protect\alpha \right\vert =\left\vert \protect\beta \right\vert
=50000$.}
\end{figure}

Moreover, to show the effect of the coherence amplitude $\left\vert \alpha
\right\vert $ and $\left\vert \beta \right\vert $ on the incident angle
sensitivity, at $\theta =43.6208%
{{}^\circ}%
$ and $\theta =43.6407%
{{}^\circ}%
$, we also plot the incident angle sensitivity as a function of $\left\vert
\alpha \right\vert $ and $\left\vert \beta \right\vert $ in Fig. 8(a) and
8(b), respectively. It can be found that, the incident angle sensitivity $%
\Delta \theta $ decreases sharply with the increase of $\left\vert \alpha
\right\vert $ and $\left\vert \beta \right\vert $, for the case of $\theta
=43.6208%
{{}^\circ}%
$ and $\theta =43.6407%
{{}^\circ}%
$.

\begin{figure}[tbp]
\label{Fig8} \centering \includegraphics[width=0.75\columnwidth]{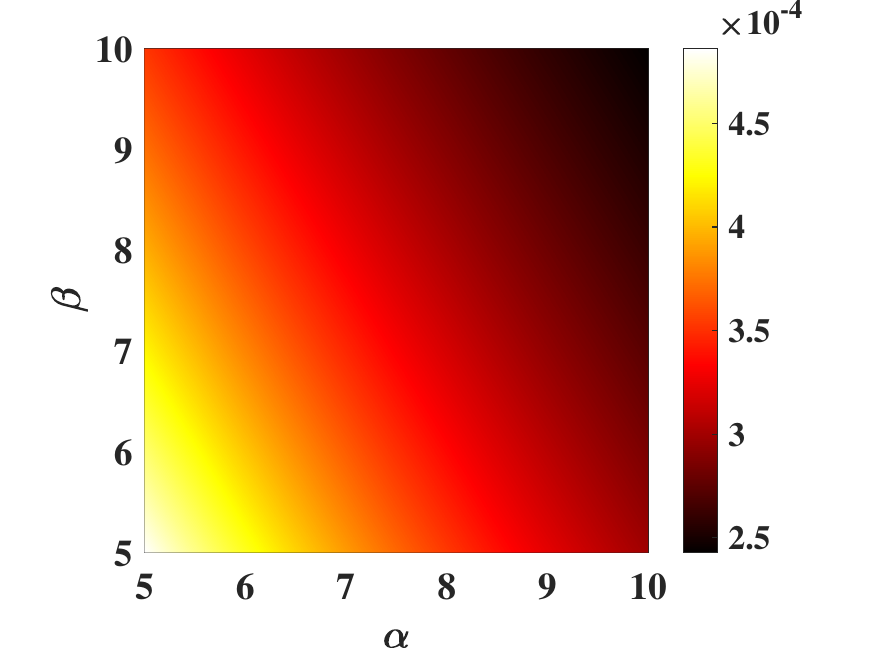}%
\newline
\includegraphics[width=0.75\columnwidth]{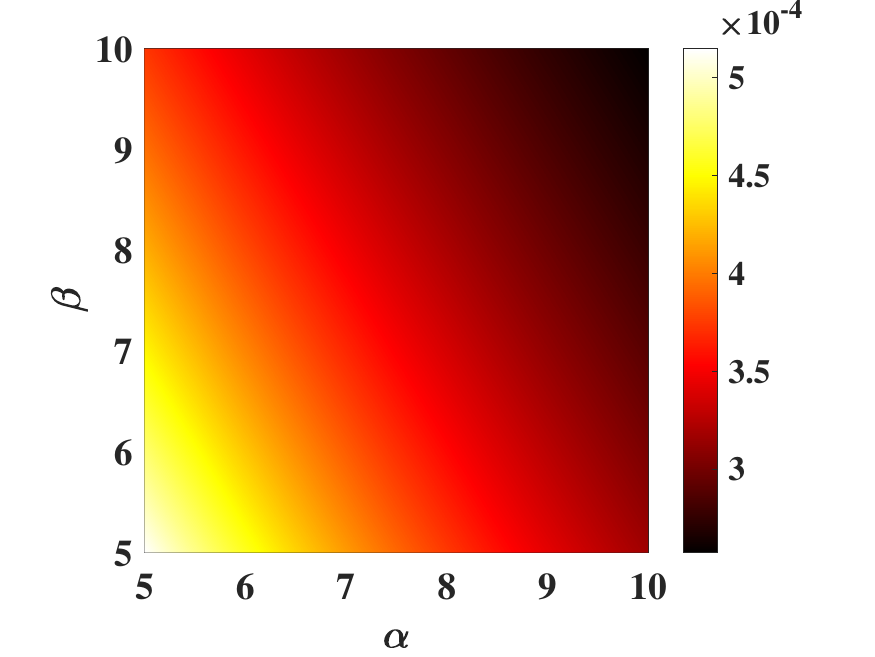} \newline
\caption{(Color online) Incident angle sensitivity based on homodyne
detection as a function of the coherent amplitude $\left\vert \protect\alpha %
\right\vert $ and $\left\vert \protect\beta \right\vert $ with $g=0.7$, $%
\protect\theta _{\protect\alpha }=0$, $\protect\theta _{\protect\beta }=%
\protect\pi $ at (a) $\protect\theta =43.6208%
{{}^\circ}%
$, and (b) $\protect\theta =43.6407%
{{}^\circ}%
$, respectively. }
\end{figure}

\section{The QCRB and SNL of IF shifts and incident angle}

\bigskip As we all know, one can also obtain the precision limit of IF
shifts and incident angle, respectively, without using any the detection
method by calculating the QFI of the probe state, which characterizes the
maximum amount of information about IF shifts and incident angle. Under
lossless cases, for a pure quantum state, the QFI of IF shifts and incident
angle can be respectively expressed as%
\begin{eqnarray}
F_{Y} &=&4[\left \langle \psi _{Y}^{\prime }|\psi _{Y}^{\prime }\right
\rangle -|\left \langle \psi _{Y}^{\prime }|\psi _{Y}\right \rangle |^{2}],
\notag \\
F_{\theta } &=&4[\left \langle \psi _{\theta }^{\prime }|\psi _{\theta
}^{\prime }\right \rangle -|\left \langle \psi _{\theta }^{\prime }|\psi
_{\theta }\right \rangle |^{2}],  \label{13}
\end{eqnarray}%
where $\left \vert \psi _{\Gamma }\right \rangle =\hat{U}_{BS}\hat{S}(\xi
_{1})\left \vert \psi _{in}\right \rangle \left \vert 0\right \rangle
_{v_{a}},$ $\Gamma =Y,\theta $ is the state vector prior to the second OPA
and $\left \vert \psi _{\Gamma }^{\prime }\right \rangle =\partial
\left
\vert \psi _{\Gamma }\right \rangle /\partial \Gamma .$ Then, for our
scheme, the QFI of IF shifts and incident angle can be respectively given by

\begin{eqnarray}
F_{Y} &=&4\left[ \left \langle \hat{H}_{Y}^{2}\right \rangle -\left \langle
\hat{H}_{Y}\right \rangle ^{2}\right]  \notag \\
&=&\frac{(d\eta /dY)^{2}}{\eta \left( 1-\eta \right) }\left \langle \hat{a}%
^{\dagger }\hat{a}\right \rangle ,  \notag \\
F_{\theta } &=&4\left[ \left \langle \hat{H}_{\theta }^{2}\right \rangle
-\left \langle \hat{H}_{\theta }\right \rangle ^{2}\right]  \notag \\
&=&\frac{(d\eta /d\theta )^{2}}{\eta \left( 1-\eta \right) }\left \langle
\hat{a}^{\dagger }\hat{a}\right \rangle ,  \label{14}
\end{eqnarray}%
where $\left \langle \cdot \right \rangle $ is the average value under the
state $\hat{S}(\xi _{1})\left \vert \alpha \right \rangle _{a}\left \vert
\beta \right \rangle _{b}\left \vert 0\right \rangle _{v_{a}}$ and

\begin{eqnarray}
\hat{H}_{Y} &=&i(d\hat{U}_{BS}^{\dagger }/d\eta )(d\eta /dY)\hat{U}_{BS},
\notag \\
\hat{H}_{\theta } &=&i(d\hat{U}_{BS}^{\dagger }/d\eta )(d\eta /d\theta )\hat{%
U}_{BS},  \notag \\
\left \langle \hat{a}^{\dagger }\hat{a}\right \rangle &=&(\left \vert \alpha
\right \vert \cosh g+\left \vert \beta \right \vert \sinh g)^{2}+\sinh ^{2}g.
\label{15}
\end{eqnarray}%
According to the Eq. (\ref{15}), one can respectively get the QCRB $\Delta
Y_{QCRB}$ and $\Delta \theta _{QCRB}$ of IF shifts and incident angle, which
denotes the lower bound of the IF shifts sensitivity, i.e.,
\begin{eqnarray}
\Delta Y_{QCRB} &=&\frac{1}{\sqrt{vF_{Y}}},  \notag \\
\Delta \theta _{QCRB} &=&\frac{1}{\sqrt{vF_{\theta }}},  \label{16}
\end{eqnarray}%
where $v$ is the number of trials and we set $v=1$ for simplicity$.$We can
clearly see from Eq. (\ref{17}) that the larger the value of $F_{Y},$ the
higher the IF shifts and incident angle sensitivity. The corresponding SNL
of IF shifts and incident angle can be respectively derived as%
\begin{eqnarray}
\Delta Y_{SNL} &=&\frac{1}{\sqrt{(d\eta /dY)^{2}\left \langle \hat{N}\right
\rangle /4\eta \left( 1-\eta \right) }},  \notag \\
\Delta \theta _{SNL} &=&\frac{1}{\sqrt{(d\eta /d\theta )^{2}\left \langle
\hat{N}\right \rangle /4\eta \left( 1-\eta \right) }},  \label{17}
\end{eqnarray}%
where $\left \langle \hat{N}\right \rangle =(\left \vert \alpha \right \vert
^{2}+\left \vert \beta \right \vert ^{2})\cosh 2g+2\left \vert \alpha
\right
\vert \left \vert \beta \right \vert \sinh 2g+2\sinh ^{2}g$ is the
total mean photon number inside the SU(1,1) interferometer.

In order to see whether the precision of IF shifts can surpass the SNL, we
make a comparison about IF shifts sensitivity, the SNL $\Delta Y_{SNL}$ and
the $\Delta Y_{QCRB}$, as shown in Fig. 9. The parameters here are the same
as those in Fig. 4. It can be seen from Fig. 9 that IF shifts sensitivity
(blue-dashed line) can break through SNL (red-solid line) and is always
surpassed by the QCRB (red-solid line). However, $\Delta Y$ gradually
approches to the ultimate IF shift precision limit $\Delta Y_{QCRB}$ as the
incident angle gets close to $\theta =43.6208%
{{}^\circ}%
$ and $\theta =43.6407%
{{}^\circ}%
$.

\begin{figure}[tbp]
\label{Fig9} \centering \includegraphics[width=0.75\columnwidth]{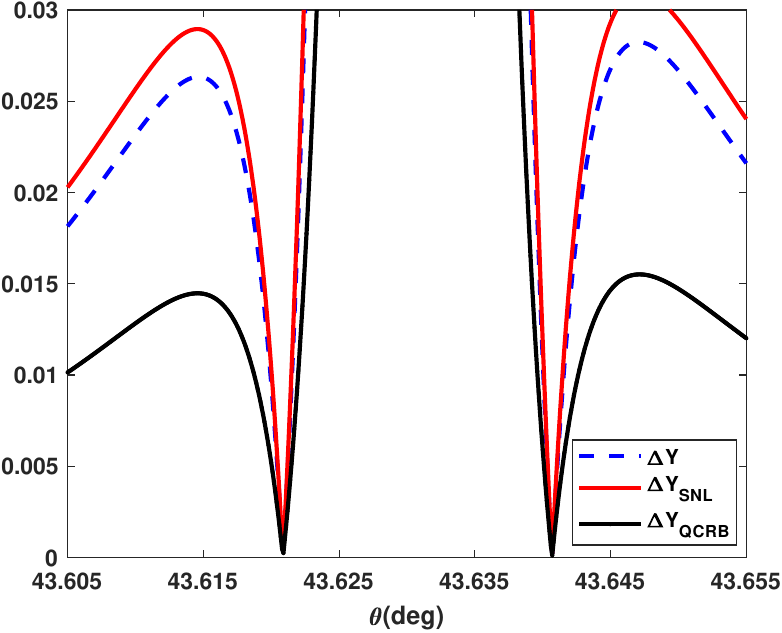}%
\newline
\caption{(Color online) $\Delta Y$, $\Delta Y_{QCRB}$ and $\Delta Y_{SNL}$
as a function of incident angle $\protect\theta $ with $g=0.7$, $\protect%
\theta _{\protect\alpha }=0$, $\protect\theta _{\protect\beta }=\protect\pi $
and $L=1$, respectively. }
\end{figure}
Moreover, to show the effects of the coherence amplitude $\left\vert \alpha
\right\vert $ and $\left\vert \beta \right\vert $ on the QCRB, we also plot
the $\Delta Y_{QCRB}$ as a function of $\left\vert \alpha \right\vert $ and $%
\left\vert \beta \right\vert $ at $\theta =43.6208%
{{}^\circ}%
$ and $\theta =43.6407%
{{}^\circ}%
$ in Figs. 10(a) and 10(b). It can be seen that the value of $\Delta
Y_{QCRB} $ decreases with the increase of $\left\vert \alpha \right\vert $
and $\left\vert \beta \right\vert $.

\begin{figure}[tbp]
\label{Fig10} \centering \includegraphics[width=0.75%
\columnwidth]{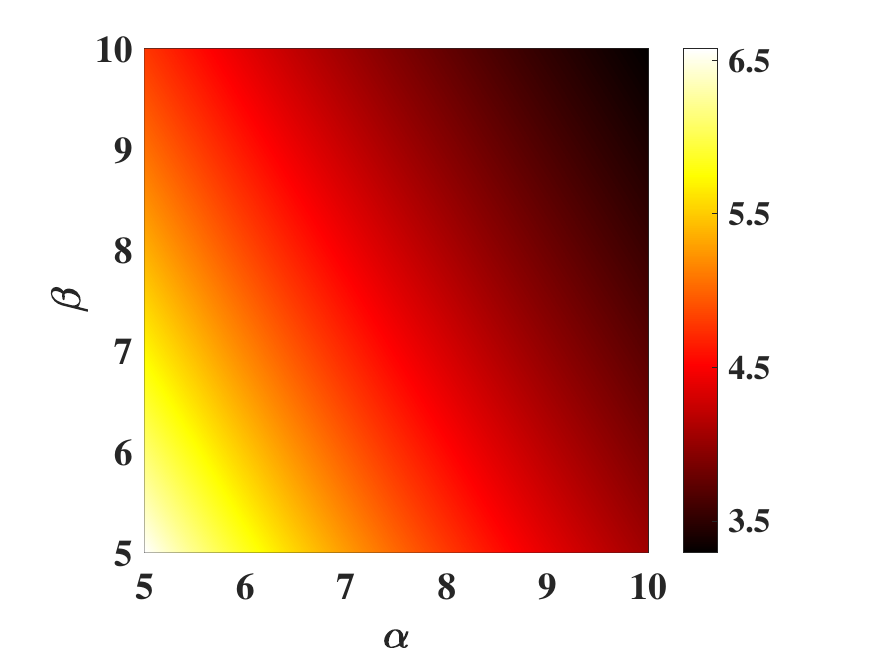}\newline
\includegraphics[width=0.75\columnwidth]{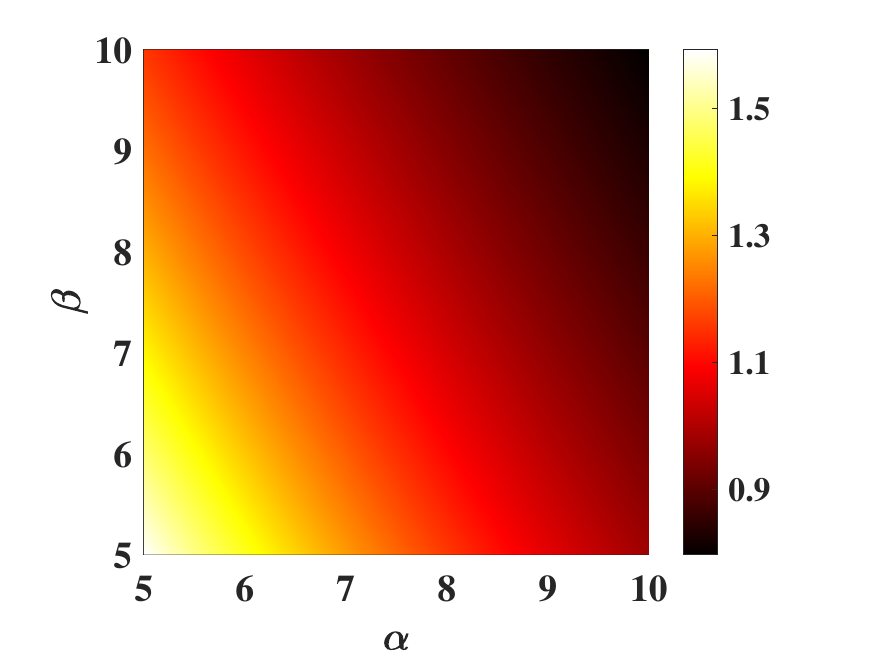} \newline
\caption{(Color online) The $\Delta Y_{QCRB}$ as a function of the coherent
amplitude $\left\vert \protect\alpha \right\vert $ and $\left\vert \protect%
\beta \right\vert $ with $g=0.7$, $\protect\theta _{\protect\alpha }=0$, $%
\protect\theta _{\protect\beta }=\protect\pi $ and $L=1$ at (a) $\protect%
\theta =43.6208%
{{}^\circ}%
$, and (b) $\protect\theta =43.6407%
{{}^\circ}%
$, respectively. }
\end{figure}
Further, we also illustrate the incident angle sensitivity $\Delta \theta $
, the SNL $\Delta \theta _{SNL}$ and the QCRB $\Delta \theta _{QCRB}$ as a
function of the incident angle $\theta $ in Fig. 11. The parameters here are
the same as those in Fig. 7. It can be seen from Fig. 11 that the incident
angle sensitivity (blue-dashed line) can also beat the SNL (green-dashed
line), but can not break through the QCRB (black-dashed line).

\begin{figure}[tbp]
\label{Fig11} \centering \includegraphics[width=0.75\columnwidth]{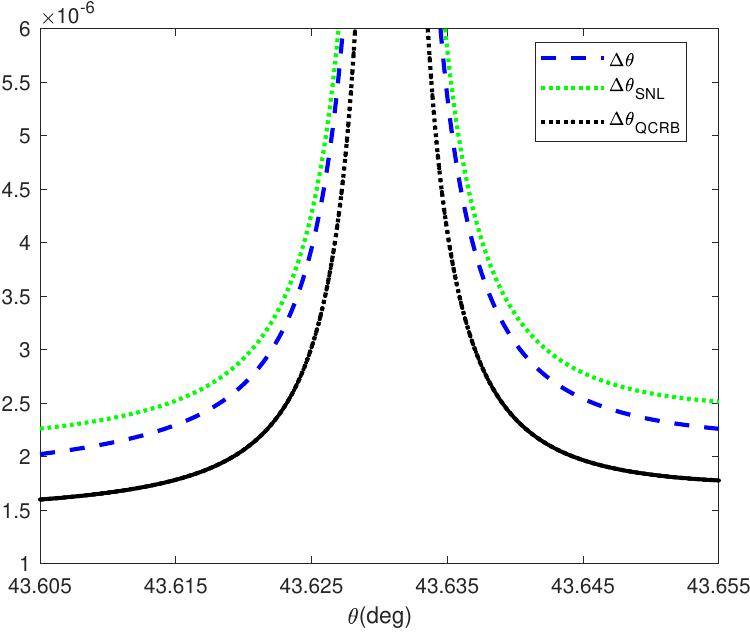}%
\newline
\caption{(Color online) $\Delta \protect\theta $, $\Delta \protect\theta %
_{QCRB}$ and $\Delta \protect\theta _{SNL}$ as a function of incident angle $%
\protect\theta $ with $g=0.7$, $\protect\theta _{\protect\alpha }=0$, $%
\protect\theta _{\protect\beta }=\protect\pi $, respectively. }
\end{figure}

\section{\protect\bigskip Conclusions}

In conclusion, we presented a theoretical estimation scheme of the IF shift
and incident angle based on the Kretschmann structure which is placed in an
SU(1,1) interferometer. Then, we respectively derived the sensitivity of IF
shifts and incident angle via the homodyne detection. Our analysis shows
that, at the incident angle $\theta =43.6208%
{{}^\circ}%
$ and $\theta =43.6407%
{{}^\circ}%
$, the maximal IF shifts and the minimum IF shifts sensitivity can be
obtained simultaneously. Moreover, we find that the quanta number of the
orbit angular momentum is unfavorable for improving the IF shift
sensitivity. Actually, the sensitivity of IF shifts can be obtained by the
weak measurement scheme. Therefore, we also investigated the incident angle
sensitivity in our scheme. The numerical results showed that the incident
angle sensitivitiy can break through $(6\times 10^{-6})%
{{}^\circ}%
$ which shows more accurate compared with the rotation precision of 0.04$%
{{}^\circ}%
$ in the weak measurement of IF shifts \cite{39}. More importantly, both the
sensitivity of IF shifts and incident angle can beat the SNL, even
approaching the QCRB at the incident angle $\theta =43.6208%
{{}^\circ}%
$ and $\theta =43.6407%
{{}^\circ}%
$. Finally, we discussed the effects of the coherent amplitude on the IF
shifts, incident angle sensitivity and QCRB. The results revealed that the
increase of coherent amplitude is beneficial to improve the sensitivity of
IF shifts and incident angle. Our work provided a novel scheme for measuring
micro-displacement in SPR sensor, which has great application potential in
the field of quantum sensor and quantum information processing.

\end{document}